\documentclass{rspublic}
\usepackage{graphicx}
\usepackage{epstopdf}

\begin{document}

\title[On the UHECRs acceleration]{On the acceleration of\\ Ultra-High-Energy Cosmic Rays}

\author[F. Fraschetti]{Federico Fraschetti$^{1,2}$}

\affiliation{$^1$LUTh, Observatoire de Paris, CNRS UMR 8102, Universit\'e Paris Diderot,\\
5 Place Jules Janssen, F-92195 Meudon C\'edex, France;\\
$^2$Laboratoire AIM, CEA/DSM - CNRS - Universit\'e Paris Diderot,\\
IRFU/SAp, F-91191 Gif sur Yvette C\'edex, France}

\label{firstpage}

\maketitle

\begin{abstract}{Cosmic Rays; High Energy Astrophysics; Acceleration of Particles; Magnetic Turbulence}
Ultra High Energy Cosmic Rays (UHECRs) hit the Earth's atmosphere with energies exceeding $10^{18}$ eV. This is the same energy as carried by a tennis ball moving at 100 km/h, but concentrated on a sub-atomic particle.
UHECRs are so rare (the flux of particles with $E > 10^{20}$ eV is $0.5$/km$^2$/century) that only a few such particles have been detected over the past 50 years.

Recently, the HiRes and Auger experiments have reported the discovery of a high-energy cut-off in the UHECR spectrum, and Auger has found an apparent clustering of the highest energy events towards nearby active galactic nuclei. Consensus is building that the highest energy particles are accelerated within the radio-bright lobes of these objects, but it remains unclear how this actually happens, and whether the cut-off is due to propagation effects or reflects an intrinsically physical limitation of the acceleration process. 

The low event statistics presently allows for many different plausible models; nevertheless observations are beginning to impose strong constraints on them. These observations have also motivated suggestions that new physics may be implicated. We present a review of the key theoretical and observational issues related to the processes of propagation and acceleration of UHECRs and proposed solutions.
\end{abstract}

\section{Introduction}

Hess (1912) discovered cosmic rays by measuring ionisation at increasing altitudes in a balloon flight. Auger et al. (1938) proved the existence of extended air showers of secondary particles caused by the interaction of primary particles, having energies above $10^{15}$ eV ($1$eV $= 1.602\times10^{-19}$J), with the Earth's atmosphere. Auger simultaneously observed the arrival of secondary particles in ground detectors set many meters apart. After almost 90 years of research, the origin of cosmic rays and the physical mechanism by which they are accelerated has remained an open question.

The Universe is filled with the remnant radiation emitted by the Big Bang, the Cosmic Microwave Background (CMB) radiation. Since we do not observe degradation of Ultra High Energy Cosmic Rays (UHECRs) due to the interaction with CMB, their sources must be located within 75 Mpc ($1$ Mpc $=10^6$ pc and $1$ pc $\sim 3.09 \times 10^{18}$ cm). 

The supernova remnant, i.e. the material ejected in the interstellar space by the explosion of a massive star, is believed to be a strong candidate for the source of medium-energy cosmic rays, although direct evidence for this is still lacking. Moreover, the processes of acceleration of particles in the supernova remnant cannot overcome the threshold of $\sim 10^{15}$eV since higher energy particles cannot be confined in such objects (Lagage \& Cesarsky 1983). Therefore the sources of UHECRs must be located elsewhere, very likely in nearby extragalactic sources.

This review is intended to cover theoretical and observational aspects of the acceleration mechanisms which lead to Ultra High Energy (UHE), without entering the mathematical details. Inevitably not all works have been properly discussed or mentioned; for a more complete review see for instance Torres  \& Anchordoqui (2004) and Bhattacharjee \& Sigl (2000).

\section{Review of observations of UHECRs}\label{obs}

\begin{figure}[htbp]
\begin{center}
\includegraphics[width=6.5cm]{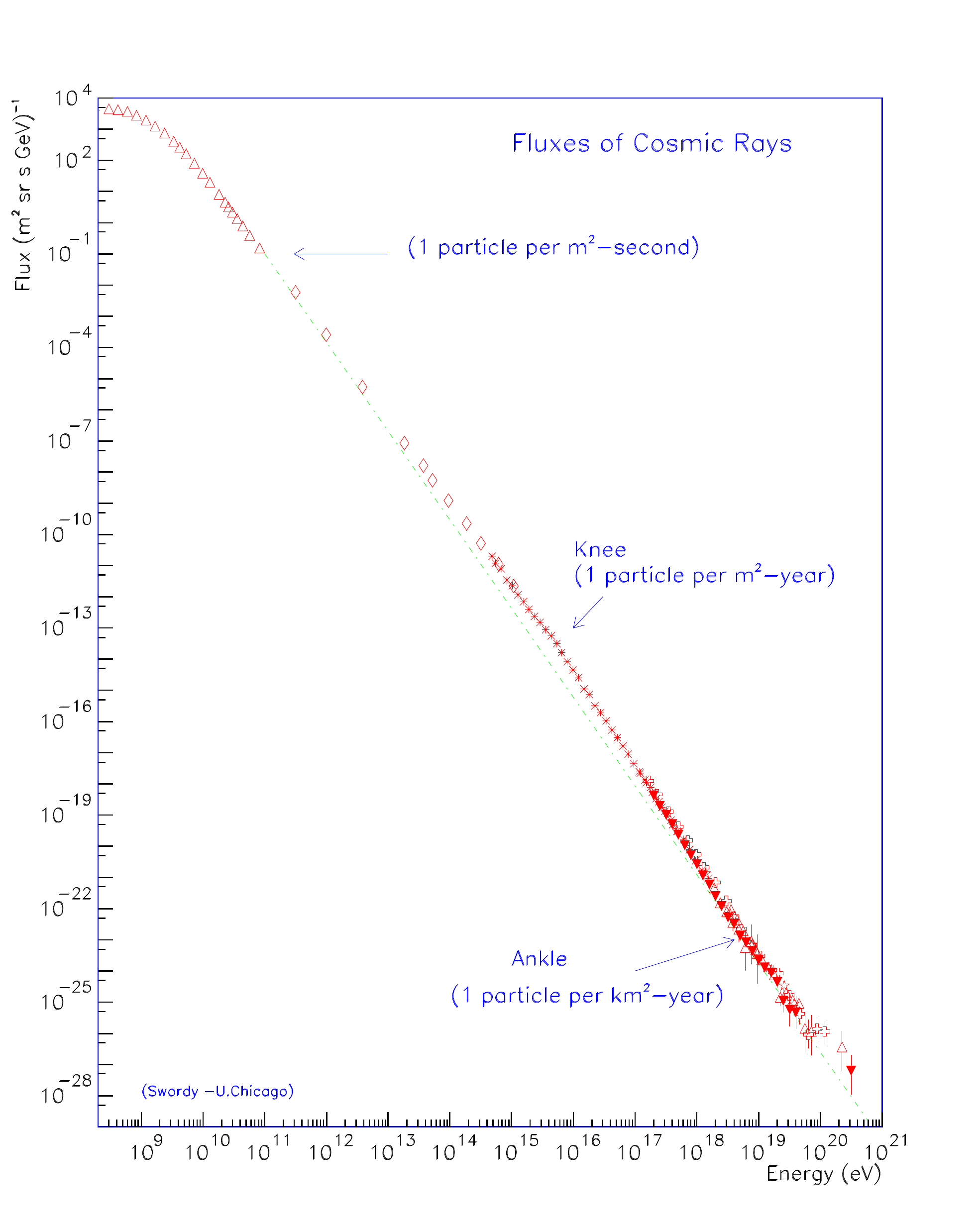}
\includegraphics[width=10cm]{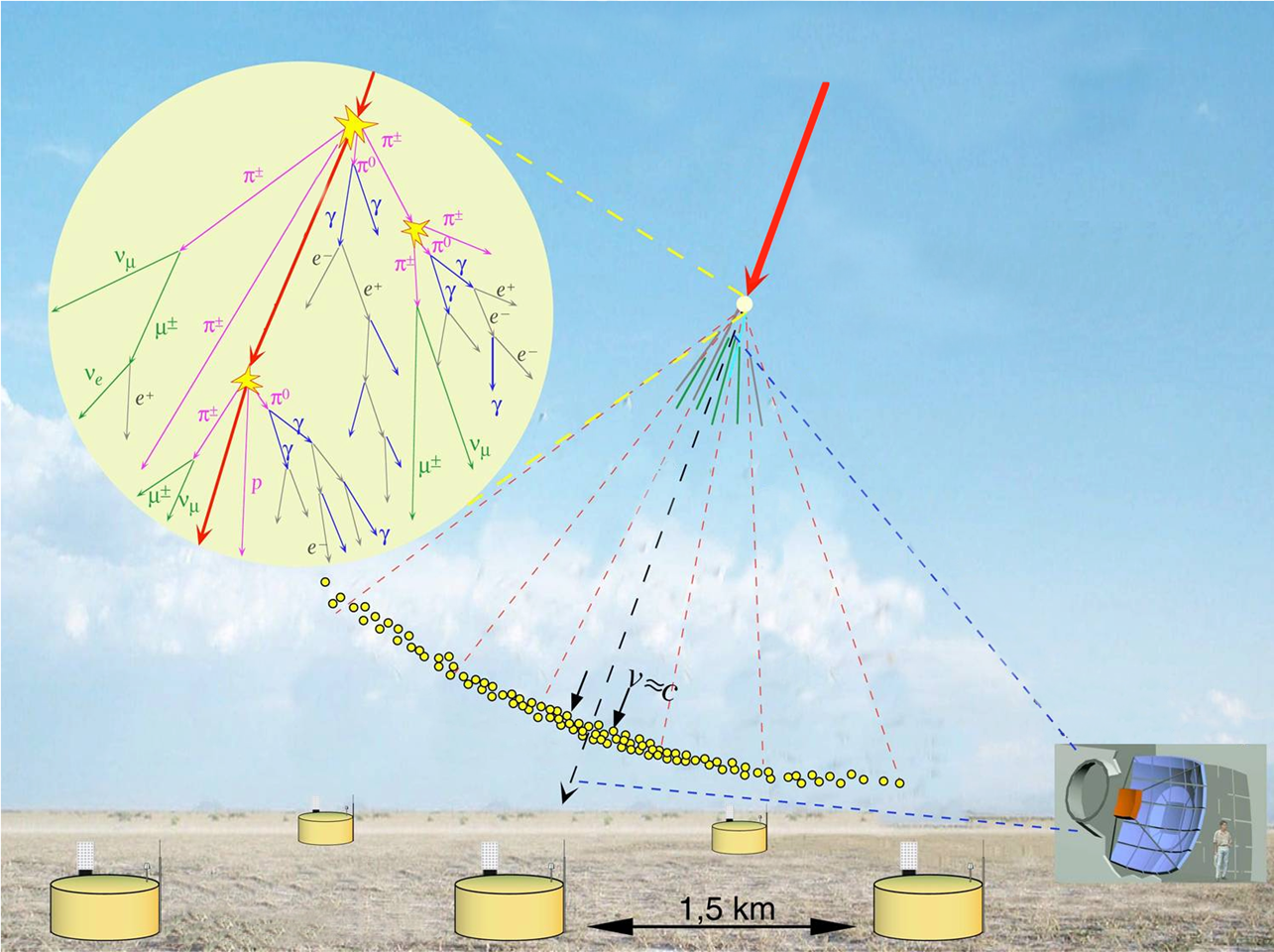}
\caption{({\it a}) The overall cosmic ray spectrum is shown (available on the URL: http://apcauger.in2p3.fr/Public/Presentation). It appears to be a unique power law with at least two changes in slope, at the ``knee'' and at the ``ankle''. The origin of the knee is still an open issue; it may result from a change in the propagation properties of particles at this energy in the intergalactic medium or it may be the signature of sources different from supernova remnants. As regards the ankle, it is clear that the UHECRs above this energy have an extragalactic origin. ({\it b}) When a cosmic ray particle collides with the nuclei in the Earth's atmosphere, it produces several secondary particles. The secondary particles too collide with other nuclei in the atmosphere giving rise to an extended shower of billions of other particles which reach the ground level over a large area, whose size depends on the energy of the primary cosmic ray (available on the URL: http://apcauger.in2p3.fr/Public/Presentation).}
\label{spect}
\end{center}
\end{figure}

The cosmic ray spectrum extends over 13 orders of magnitude in energy and more than 30 orders of magnitude in observed flux (see figure~\ref{spect}{\it a}). The cosmic ray flux at energy $E > 10^{20}$eV is estimated to be about $0.5$ particle/km$^2/$century. Qualitatively, the current experiments agree that the overall cosmic rays spectrum can be divided into the following four regions: 1) below $10^9$ eV it is flat and the particles are decelerated by the magnetized plasma ejected by the Sun's surface; 2) between $10^{10}$ eV and $3\times 10^{15}$eV (the so-called ``knee'') it can be fitted by a unique power law with index $2.7$ and is likely to be dominated by galactic supernova remnants; 3) between $3\times 10^{15}$eV and $4\times 10^{18}$eV (the so-called ``ankle'') the galactic contribution of primary cosmic rays heavier than protons, such as He or Fe, may be boosted with acceleration in other environments, such as pulsar winds, namely wind of relativistic particles emitted by a rapidly rotating and pulsating star; 4) at energies $E > 4\times 10^{18}$eV, in the UHE region, some experiments have provided indication for a slope around $2.6$, and the steepening of the galactic component would progressively be drowned by a flatter extra-galactic flux. In the knee region, KASCADE data have confirmed (2008) an overlap of heavier elements, such as Fe. A similarly non negligible contribution of heavy nuclei has been found in UHE region by The Auger Collaboration (2008$a$).

At an energy greater than $10^{14}$ eV, the cosmic rays impinging on the Earth's atmosphere produce extended showers; from a primary cosmic ray, the shower spreads into a number of detectable secondary particles over a diameter of 100 meters. At $10^{20}$eV this diameter has increased to 5 km (see figure 1 {\it b}). From the particle densities measured in the detectors, the energy of the extended shower, and therefore of the primary cosmic ray, can be inferred. The main difficulty in the analysis of such observations is that the energy of the collision of primaries with the Earth's atmosphere (in the center of mass frame) is well beyond the limits of the particle accelerators, even the Large Hadron Collider (LHC) at CERN (acronym standing in French for Conseil Europ\'een pour la Recherche Nucl\'eaire). Hence the reconstruction of the showers relies on phenomenological extrapolation of models of particle interactions, and uncertainties on the nature of the primaries grow with the energy. In the near future, new measurements at the LHC will perhaps help the development of models of these interactions.

In the last twenty years or so, observations have been made by the following experiments: Haverah Park in UK, Akeno-AGASA in Japan, covering 100 km$^2$, Yakutsk in north-east Siberia, High Resolution Fly's Eye, in Dugway, Utah, and Auger South in western Argentina, spread over an area of 3000 km$^2$, similar in size to Rhode Island.

Haverah Park and Akeno measure the density of the secondary particles in the shower at ground level at a distance of 0.6 km from the core. The Yakutsk experiment uses this density technique but calibrated by an array of wide angle photomultiplier detectors that observe the Cerenkov light, namely the radiation emitted by particles propagating through a medium at a velocity greater than the velocity of light in that medium. HiRes measures the energy from the fluorescence of the showers in the atmosphere, by observing the flash of nitrogen molecules which de-excite after the passage of the shower. The Auger South Observatory (see figure~\ref{spect}, {\it b}) uses both these complementary techniques to measure the extended showers on the ground with 1600 detectors spaced 1.5 km apart and in the atmosphere using four fluorescence telescope arrays. The latest observations by The Auger Collaboration (2007) have shown that the sources of cosmic rays with energy $E > 6 \times 10^{19}$eV are correlated with the matter distribution within 75 Mpc, specifically Active Galactic Nuclei (AGN). This correlation does not identify unequivocally the nearby AGNs as the sources, but rather indicates that any astrophysical class of objects with a local spatial distribution sufficiently similar to that of AGNs is a possible accelerator of UHECRs. 

The GZK cut-off (Greisen 1966, Zatsepin \& Kuz'min 1966) is the upper limit on the energy that a cosmic ray proton can retain while propagating through the CMB. This should be evident in the cosmic ray energy spectrum if the sources of UHECRs are homogeneously distributed through the Universe. Recent observations by The Auger Collaboration (2008{\it b}) The HiRes Collaboration (2008)HiRes (2008) and Auger (2008$b$) have reported the discovery of a cut-off at UHE in the cosmic ray spectrum. Whether the spectral steepening is a manifestation of the long-sought GZK cut-off or an intrisic limitation of the acceleration process, such as the geometry or the nature of the magnetic turbulence of the source, still needs to be understood. Moreover, difficulties in the definition of the GZK cut-off lie in the model dependence of the shape of the cosmic ray spectrum beyond the ankle: as an example, a local overdensity of UHECR sources, assumed to be isotropically distributed within a distance of a few tens of Mpc, drastically changes the spectrum in the GZK cut-off region (Berezinski \textit{et al.} 2006). A precise definition of the GZK cut-off is also affected by the fact that the details of the spectral shape in the UHE range depend on the nature of the primaries, as the cut-off of heavy nuclei such as Fe is expected at lower energies than that of protons.

\section{Propagation of high energy particles}\label{transport}

Beyond the ankle ($E \sim 4 \times 10^{18}$ eV), the energy of a microscopic particle becomes macroscopic, i.e.  $10^{20}$eV is the same amount of energy as carried by a tennis ball moving at 100 km $h^{-1}$. Thus the existence of such high-energy particles, in itself, represents a major theoretical challenge and an explanation of the mechanism of acceleration based on first principles needs to be found.

The length scale characterizing the motion of a particle of energy $E$, electric charge $Ze$ in a magnetic field $B$ is the gyration radius $r_g$, or gyroradius: $r_g = E/(ZeB)$; a greater gyroradius implies a less curved trajectory. If the UHE is attained through an acceleration process, a general estimate of the maximal energy can be obtained by requiring the gyroradius $r_g$ of the particle to be no larger than the linear size $R$ of the accelerator:
\begin{equation}
r_g \sim 10^{2} E/(Z B) < R\quad;
\label{rg}
\end{equation}
if $B$ is measured in $10^{-10}$ T, $E$ in $10^{20}$ eV, then $r_g$ is measured in kpc. Equation (\ref{rg}) is also called the Hillas criterion (Hillas (1984)). It neglects the finite lifetime of the acceleration region and energy losses due to interactions with the environment such as synchrotron radiation in the magnetic field and production of secondary particles. A more general expression for the maximal energy is
\begin{equation}
E_{max} \sim \alpha 10^{18} Z R B \quad eV
\label{Emax}
\end{equation}
with $\alpha = \mathcal{E}/B$, where $\mathcal{E}$ and $B$ are respectively the electric and magnetic fields. The formula (\ref{Emax}) can be derived from dimensional arguments, representing the electromotive force associated with the time variation of magnetic field. Charged particles are accelerated by the electric field; however cosmic plasmas do not allow high voltages to be maintained. With regard to cosmic rays at $10^{20}$eV, there are no known astrophysical environments where voltages high enough can be generated. However, time variation of a strong magnetic field can generate electromotive forces which through a strong induced electric field may lead to extremely high energies. 

\begin{figure}[htbp]
\begin{center}
\includegraphics[width=7.5cm]{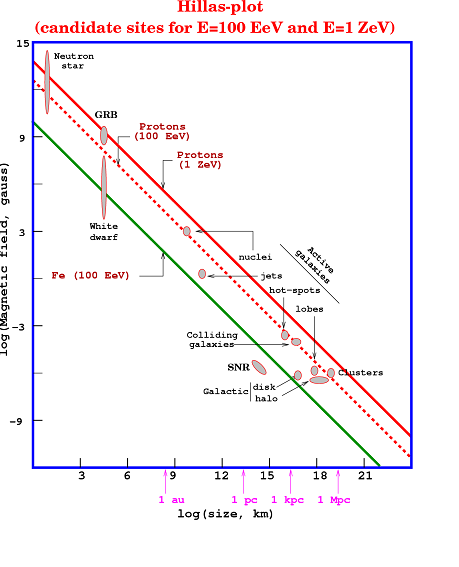}
\caption{The Hillas plot (candidate sites for $E = 100E$ eV and $E = 1Z$ eV, where $1E$ eV $=10^{18}$ eV and $1Z$ eV $=10^{21}$ eV) showing the linear size $R$ versus the magnetic field $B$ of the region able to accelerate up to $E \sim 10^{20}$eV (available on the URL: http://www.pi1.physik.uni-erlangen.de/$\sim$kappes/lehre/WS05-VAT/V9/Hillas-plot.png). The diagonals lines show the regions for which the product $BR$ satisfies the criterion 2 for different composition of the primary cosmic rays, i.e. protons or iron nuclei. As can be seen, supernova remnants (SNR) are excluded.}
\label{hillas}
\end{center}
\end{figure}

Even if only qualitative, equation (\ref{Emax}) provides an interesting criterion to identify possible sources of UHECR by simply looking at the largest values of the product $BR$. It is remarkable (see figure~\ref{hillas}) that the possible accelerators range from Neutron stars ($10$ km), namely remnants that can result from the gravitational collapse of a massive star in a supernova explosion, up to galaxy clusters ($1$ Mpc).

We should keep in mind that at a given energy the gyroradius is larger for smaller charge of the particle. Therefore,
if UHECRs are mostly protons, due to their large gyration radius, they are not deviated significantly by magnetic fields, so they should point back to their sources within an angle that depends on the intensity of the intergalactic magnetic field. For heavy nuclei, the effect of the magnetic field becomes more important.

Theoretical models can be divided in two classes: the so-called ''bottom-up'', in which a specific process of acceleration in a particular astrophysical object leads to UHEs (see the review by Torres \& Anchordoqui 2004); and the so-called ''top-down'', where UHE particles are produced by the decay of super-heavy dark matter particles or collision among cosmic strings or topological defects (see the review by Bhattacharjee \& Sigl 2000).

\section{Astrophysical models}\label{astromod}

In this section we give a short presentation of some of the most promising astrophysical models of acceleration to UHE. Owing to the low statistics involved it is difficult to favour one model over another: many of these models are equally plausible. Only upcoming observations will allow us to discriminate between them. These models may be  divided in two classes according to the acceleration process, either magnetohydrodynamical ($a$), MHD, or stochastic ($b,c,d$).

\subsection{Non-stochastic models}

UHE can be attained through the drift of charged particles in magnetized, cylindrically collimated and ultrarelativistic flows of plasma emitted by active galaxies (Lyutikov \& Ouyed 2007). If the velocity of the flow increases towards the axis of the jet, there will be a radial induced electric field which can accelerate particles. The acceleration efficiency is inversely proportional to the charge, thus protons are expected to dominate over heavy nuclei in UHECR composition. In order for the particles to reach an energy of $3\times 10^{20}$eV, a minimal luminosity $L = 10^{39}$ W is required; therefore only very high power radio-galaxies are the favourite accelerator candidates. This mechanism needs a seed population of pre-accelerated particles at energy $\lesssim 10^{18}$ eV at the base of the centre of the galaxy or even outside the jet. 

Recently it has been proposed that supermassive black holes (with mass $M \sim 10^8 M_{\odot}$, where $M_{\odot} = 1.98\times10^{30}$ kg is the mass of the Sun) might also be accelera\-tors up to UHE (Neronov \textit{et al.} 2007). If the axis of magnetic field is misaligned with the rotation axis of the black hole, charged particles, initially spiralling into the black hole along the equatorial plane, penetrate into polar regions of the black hole and are ejected to infinity with very high energies. It is a necessary condition in this scenario that the magnetic field is almost aligned to the rotation axis of the black hole.

It has also been speculated (Boldt \& Ghosh 1999) that presently radio-quiet quasars, containing black holes with mass $M > 10^9 M_{\odot}$ in the centre of galaxies located within a distance of $100$ Mpc, could be the accelerators to UHE. An external magnetic field, generated by the external currents flowing in the surrounding disc, threads the black hole; thus, an electromotive force is induced. The loss of energy of the accelerated particles would allow only a limited number of particles to reach energies $E> 10^{20}$eV, but this model is not yet falsified by observations.

Young neutron stars with surface magnetic field in the range $10^{8}-10^{10}$ T could accelerate iron nuclei through relativistic hydrodynamical winds in a magnetic field up to energy $10^{20}$eV (Blasi \textit{et al.} 2000). The iron nuclei can pass through the surrounding supernova remnant without suffering significant deceleration and thus be observed. However, the discovery of the high-energy cut-off disfavours this model which predicts observations above the cut-off.

\subsection{Fermi second-order mechanism}\label{fermi}

The original idea of explaining the acceleration of high energy cosmic rays by stochastic collisions is due to Fermi (1949) who imagined that a particle can be accelerated through random collisions with the irregularities of the interstellar ma\-gnetic field. The energy source is therefore the time variation of the intergalactic magnetic field which, through the induced electric field, accelerates particles. This is the only possible origin of acceleration because a constant magnetic field, with any possible orientation, only affects the trajectory but does not do any work on the particle. On the other hand, in intergalactic space any background electric field vanishes, because the ionized gas in the interstellar medium is highly conductive and therefore globally neutral.

As shown by Fermi, the average energy gain is given by $\langle \Delta E / E \rangle \sim 4/3 \times \beta ^2 $, where $\beta$ is the mean non-relativistic velocity of the magnetic irregularities in units of the velocity of light in vacuum, $c=3\times 10^8$m s$^{-1}$. It is remarkable that the energy gain increases as $\beta^2$, depends only on the velocity of the clouds, and not on the magnetic field strength and/or the electric charge of the particle.

Now we consider a number of particles continuously injected in the magnetically turbulent region described previously. The second-order Fermi mechanism yields a spectral distribution $N(E)$ in the form of a power law: $N(E) \sim E^{-n}$, where $N(E)$dE is the number of particles with energy between $E$ and $E + $dE. It is remarkable that, if the spectrum has a power law dependence such as $N(E) \sim E^{-n}$ in a certain energy range, the total power emitted also has the form of a power law in the corresponding energy range. This property is at the basis of the widespread use of the power law distribution of ultrarelativistic electrons in many branches of high energy astrophysics.

However, a few issues affect the original model of Fermi. First of all, the random velocity of the magnetic clouds is of order of $\beta = 30,000$ m s$^{-1}$; thus, the average gain of energy is very small, that is $\langle\Delta E / E\rangle \sim 10^{-8}$. Since the average distance between clouds can be estimated to be of order of $1$ pc, with one collision per year the energy of a particle increases only of a factor 3 in one billion years. We know currently that the average age of cosmic rays is $10$ Myr, therefore this mechanism in its original form is too slow to attain UHEs. Second, the value of $n$ is not unequivocally predicted by the theory, since it has a complicated dependence on the propagation properties of the particles. The third issue concerns the fact that the particle injected at low energy can lose its energy in coulombian interactions; in order to avoid this, a lower limit on the initial energy was already inferred by Fermi: 200 MeV for protons, but higher for heavy nuclei as Si or Fe.

In a modern version of Fermi's mechanism, the clouds of turbulent magnetic field can be considered an idealization of the clumps of matter homogeneously distributed in the galactic space moving at Alfven wave velocity $v_A$, given by $v_A = B_0/\sqrt{4\pi \rho}$, where $B_0$ is the strength of background magnetic field and $\rho$ is the matter density of background plasma. For mercury at room temperature $v_A$ [m s$^{-1}$]$= 10^{-7}\times B_0$[T], while the sound speed is $c_s = 1.4 \times 10^3$ m s$^{-1}$. In astrophysical environments, $v_A$ can, however, be much larger owing to the low density: in the Sun's photosphere the density is of order of $10^{-4}$kg m$^{-3}$, so that $v_A$[m s$^{-1}$]$ = 10^{-3} B_0$[T]. Since in this scenario the scatterings are much more frequent than in the original Fermi model, the acceleration is much more efficient.

\subsection{Acceleration at shocks}\label{first}

More popular and efficient than Fermi second-order is the acce\-leration at the non-relativistic shock preceding the expansion of matter flowing at speeds larger than the speed of sound in that medium, e.g. the remnant of supernovae. This is the so-called first-order Fermi process (Blandford \& Eichler 1987). In this case, the average energy gain is $\langle\Delta E / E\rangle \sim 4/3 \times \beta $, where $\beta$ is roughly the shock velocity. The basic idea is that the second-order process is not very efficient because the energy gain results as the small positive balance between the gains in the head-on collisions with the clumps of matter and the losses in the overtaking collisions. By contrast, in environments such as supernova shocks the collision of a particle with a magnetically turbulent plasma, extended both upstream or downstream, is {\em always} head-on.

The first-order mechanism is more efficient than the second-order one as long as the velocity of the shock is larger than $v_A$. At late times in the supernova expansion, when the kinetic energy has been mainly dissipated in the expansion or sapped by the escape of high energy particles, the shock velocity becomes comparable to $v_A$ and the second-order effect can even dominate over the first-order one. Moreover, if the shock wave is relativistic, the first-order Fermi acceleration seems to be unable to attain ultra-high energies (see Ostrowski, in press, and references therein).

\subsection{Radio lobes of AGN}\label{paper}

As confirmed by the latest Auger observations (2007), AGN and their extensions are possible candidate sites of cosmic ray acceleration. The stochastic mechanism proposed by Fermi deserves to be explored in more detail. An attempt in this direction has been recently proposed by Fraschetti \& Melia (2008). The three-dimensional motion of individual particles is followed during their propagation within a turbulent magnetic field. The turbulence is generated by the prescription of Giacalone \& Jokipii (1994), modified to allow for temporal variations. 

In the treatment by Fraschetti \& Melia (2008), the remaining unknowns are the partition of energy between turbulent and background fields, and the turbulent energy distribution, though this may reasonably be assumed to be isotropic (Kolmogorov).

Furthermore, this acceleration mechanism is sustained over 10 orders of magnitude in particle energy, up to $E \sim 10^{20}$ eV and the UHECRs therefore emerge naturally from the physical conditions thought to be prevalent within AGN radio lobes. The theoretically predicted spectrum is in accordance with the recent observations if only the closest sources are considered (Gelmini \textit{et al.} 2007).

A new cut-off consisting of the size of the acceleration region is introduced in addition to or substitution for the GZK cut-off. This result reopens the question of whether the high-energy cut-off is due to interaction with CMB, or it  represents the maximum energy permitted by the acceleration process itself.

\section{Non-astrophysical models}\label{noastromod}

The difficulty of finding a physical mechanism of acceleration based on simple phy\-sical principles and the distance limit posed by the interaction with the CMB have lead to the development of other non-astrophysical models for the origin of UHECRs. In these models, the acceleration problem is circumvented by assuming the existence of new particles or new interactions beyond the standard model of elementary particles. Within the standard model the only particle whose path is not limited by the interaction with the CMB is the neutrino, which may have been produced as secondary particle by a primary proton accelerated to UHE in an AGN environment at a distance larger than $100$ Mpc. The neutrino-nucleon interaction in Earth's atmosphere may be enhanced by possible new physics, but this is unlikely to be able to simulate hadronic interactions (Anchordoqui \textit{et al.} 2003).

By contrast, in the so-called ``top-down'' models (Bhattacharjee \& Sigl 2000), the UHE particles are the decay product of some currently unknown supermassive ``X'' particle with mass $m_X \gg 10^{20}$eV, and energy up to $m_X$. Therefore no acceleration mechanism needs to be searched for. The X particles can be produced in two ways: if they have a short lifetime, they have to be produced continuously by certain topological defects left over from a cosmological phase transition that occurred in the early Universe and their sources are evenly distributed in the Universe. On the other hand, if they have been produced only in the early Universe, they would be distributed like the local dark matter in the halo of our Galaxy and we would observe the UHECRs coming from
their slow decays.

These non-astrophysical solutions to the UHECRs problem have the advantage of connecting with new ideas beyond the standard model of particle physics, such as grand unification and supersymmetry, as well as affording a possible probe of conditions in the early Universe. However, by their very nature these are very speculative ideas (Sarkar 2004).

In such ``top-down'' models the arrival directions of UHECRs are not expected to cluster on small angular regions of the sky, as in fact the recent Auger observations indicate. Moreover, The Auger Collaboration (2008$a$) has recently excluded all such models as possible sources of UHECRs at energy $E < 10^{20}$eV by demonstrating that the primaries are not photons as expected in such models.

\section{New perspectives}\label{sum}

The importance of elucidating the mechanism responsible for the acceleration of UHECRs is confirmed by the growing number of dedicated experiments which will join Auger South: Auger North, with an exposure $7$ times larger than the Auger South's; the JEM-EUSO mission, looking downward from the International Space Station to the dark side of the Earth; the proposed orbiting wide angle light collector (OWL), etc.
The forthcoming data on UHECRs from Auger along with the other observatories will significantly boost the statistics and drive the UHECR source identification. We should be able to investigate the nature of the spectral steepening and address the question of whether the cut-off in the cosmic ray distribution is indeed due to photo-meson interaction or the result of limitations in the acceleration process itself. 
The primary particles have been shown not to be photons, thus favouring the bottom-up scenario with respect to top-down models. 
Presently the inferences concerning the composition of UHECR rely on hadronic interaction models that extrapolate the current particle physics phenomenology up to energies well below the LHC region. It is quite likely that further investigations on the nature of UHECRs will have consequences for our understanding of the physics of fundamental interactions.

\begin{acknowledgements}
The author thanks F. Melia for useful and stimulating discussions and S. Sarkar for the revision of the manuscript. The work was supported by CNES (French Space Agency) and was carried out at CEA/Saclay and LUTh.
\end{acknowledgements}

\label{lastpage}

\end{document}